\documentclass{mem}
\usepackage{graphicx}
\usepackage{times}
\usepackage{balance}
\usepackage{natbib}
\usepackage{txfonts}
\usepackage{subfigure}
\usepackage[a4paper]{hyperref}
\idline{82}{3}

\begin{document}

\title{
Turbulence, Reconnection and Cosmic Rays in Galaxy Clusters
}

   \subtitle{}

\author{A. Lazarian\inst{1}
\and G. Brunetti\inst{2}
          }
\offprints{lazarian@astro.wisc.edu}
 
\institute{
Department of Astronomy, University of Wisconsin-Madison, 475 N. 
Charter St., Madison, WI, 53706, USA
\and
INAF-- Istituto di Radioastronomia, via Gobetti 101,
40129 Bologna, Italy
}

\authorrunning{Lazarian \& Brunetti}

\titlerunning{Turbulence Reconnection and CRs}

\abstract{
Recent years have been marked by substantial changes in our understanding 
of magnetic turbulence and magnetic reconnection, 
which, in its turn induced better understanding of cosmic ray diffusion 
and acceleration. 
Current models of magnetized 
turbulence are no more ad hoc constructions, but numerically tested theories. 
In this very short review we summarize topics presented in two talks 
given at the conference and 
provide a brief sketch of the vast and rapidly developing field. 
We discuss how turbulence decreases the efficient mean free path 
of the particles in the 
collisionless plasmas in galaxy clusters
and claim that this makes MHD turbulence description
applicable to a wider range of scales. 
We discuss the properties of MHD turbulence and 
its relation to magnetic reconnection. Finally, we overview how turbulence
induces particle acceleration via second order 
Fermi process and affects first order 
Fermi acceleration in shocks and reconnection regions. 

}
							  
\maketitle{}

\section{Guide to the review}
Mergers between galaxy clusters are the most energetic events 
in the present day Universe. 
During these mergers a fraction of the gravitational energy can be converted
into fluid motions, 
i.e. shocks and turbulence, that generate magnetic fields and, 
through a variety of processes, accelerate relativistic protons and 
electrons \citep[e.g., ][]{Ryu2003, CB2005, BL2007, Hoeft2007, 
Pfrommer2008, Skillman2008, B2009, Vazza2009}.
In this short review
we address some of
the basic processes involved, namely, magnetic turbulence in 
galaxy clusters and 
the possibility of its observational studies (Sect. 2), 
magnetic reconnection (Sect. 3),
as well as various ways of accelerating cosmic rays (Sect. 4). 
Our summary is presented in Sect. 5.

\section{Turbulence in clusters of galaxies}

Astrophysical fluids are characterized by high Reynolds numbers and are 
known to be turbulent \citep[e.g., ][]{Armstrong1995,
ChL2010, Elmegreen2004, McKee2007}.
As properties of turbulent magnetized fluids are very different from 
laminar ones, 
the {\it correct} description of the particle acceleration requires 
taking into account 
the fundamental properties of magnetic turbulence 
as well as the mutual feedback of 
magnetic fields and cosmic rays in the turbulent fluids. 

\subsection{Properties of intracluster plasmas: instabilities 
induced by turbulence and effective collisions}

Turbulence in galaxy clusters is magnetized. A very important 
question is whether the MHD 
description of turbulence is 
applicable. 
When Coulomb collisions in 
the rarefied inter-galactic medium (IGM) are considered
one has to conclude that the plasma is collisionless. 
This strongly affects the proparation of compressible
modes, cosmic ray acceleration etc (see \cite{BL2007} and ref. therein).
In what follows we argue that the degree of collisionality of astrophysical 
plasmas is underestimated when only Coloumb collisions are taken 
into account (see \cite{L2010, BL2011a}).

It is well known that the mean free path of thermal protons due to 
Coulomb collisions in the hot IGM is very large, ten to hundred kpc 
\citep[e.g.,][]{Sarazin1986}.
Fluids in such a collisionless regime can be very different from their 
collisional counterparts \citep{Schekochihin2005,  Schekochihin2006, 
Schekochihin2010}.
Several instabilities (e.g. firehose, mirror,
gyroresonance etc) can be generated in 
the IGM in the presence of turbulence, leading to a 
transfer of the energy of
large-scale compressions to perturbations on smaller scales. 

Many instabilities have growth rate which peaks at scales
near the particle gyroradius,
making very large the scale separation between the energy injection scale 
and the scale where this energy is being deposited.
The scattering induced by instabilities dramatically
{\it reduces the effective mean free path} of thermal ions  
{\it decreasing the effective viscosity} of the IGM
and making plasmas {\it effectively collisional} on smaller scales. 
Indeed, charged particles can be randomized if 
they interact with perturbed magnetic field. If this field is a result 
of plasma instabilities, {\it the process can be viewed as
the collective interaction of an individual ion with the rest of the 
plasma}, which is the process mediated by magnetic field. 
As a result, the fluid would behave as collisional on scales less that 
the Coulomb mean free path. 
This issue has been addressed in \cite{LBe2006}
for the case of a collisionless fluid subject to the 
gyroresonance instability that is driven by the
anisotropy of the particle distribution in the momentum space
that arises from magnetic field compression; the larger the magnetic 
field compression, the higher the anisotropy induced and the higher is 
the instability growth rate. 
They found that the turbulent magnetic compressions on the scale of the
mean free path and less are the most effective for inducing the 
instability\footnote{The larger scale compressions do still induce the 
instability, but 
their effect is reduced due to their reduced ability to induce large 
changes of $B$ over the time scale between scattering. 
The model is further elaborated and improved
in \cite{YanL2011}. }.
As the scattering happens on magnetic 
perturbations induced by the instability, the mean free path of 
particles decreases as a result of the operation of the instability. 
This results in the process being self-regulating, i.e. the stronger 
the turbulence at the scale of injection, the smaller is the mean free 
path of plasma particles and the larger is the span of scales over 
which the fluid behaves as essentially collisional. 

This induces an interesting picture where the 
mean free path of plasma protons depends on
the level of compressions induced 
by turbulence and the mean free path is determined not 
by Columb collisions, but
scattering on magnetic field inhomogeneities at the 
Larmor radius of thermal protons. 
The peculiar feature of this picture is that the aforementioned
magnetic field perturbations are not part 
of the normal turbulent cascade, but results of 
compressible turbulent motions at much larger scales. 
Thermal protons do not scatter each
other through electric interactions, but participate 
in non-local interactions mediated by the
perturbed magnetic field. 
{\it The higher the level of comressible turbulence, the
better is MHD description of the IGM}.

\subsection{MHD turbulence: brief summary of theory and main
properties of turbulence in the IGM}

The last decade has been marked by 
substantial advances in understanding of magnetic turbulence in 
the MHD regime
\citep[e.g.,][]{GS1995, LV1999, Cho2000, Muller2000, Lithwick2001,
Cho2002, ChoL2002, ChoL2003, BeL2010, Kowal2010}.

The presence of a magnetic field makes MHD turbulence anisotropic 
\citep{Montgomery1981, Matthaeus1983, Higdon1984, Oughton2003}. The 
relative importance of hydrodynamic and magnetic 
forces changes with scale, so the 
anisotropy of MHD turbulence does too.
A landmark event in this was a seminal work by \cite{GS1995} (GS95)
which contained ideas that radically changed
the further development of the subject. The corner stone of this model was the 
so-called {\it critical balance} 
idea which provided the analytical relation between the 
fluctuations parallel and 
perpendicular to the magnetic field. 
It also contains prophetic statements about mode 
coupling, providing guidelines for generalization of the model 
from the incompressible 
to compressible MHD.

The original model was improved in the subsequent publications. 
For instance, GS95 uses the closure relations that employ
in the global system of reference related to the mean field, which, in fact, 
is an incorrect system to be used for the critical balance description. 
In \cite{LV1999} and later publications \citep[e.g.,][]{Cho2000, 
Maron2001, Cho2002}
the importance of the {\it local system of reference}, 
which is defined by the local direction of the magnetic field of 
a wave packet, was reviled. 
The local system of reference was employed in the successful 
testing of the GS95 model. In addition, \cite{LV1999}
generalized the GS95 model for the case when the turbulent 
injection velocity at the injection scale is less than the Alfvenic velocity. 

The predictions of the GS95 model are in rough agreement 
with numerical simulations \citep[e.g.,][]{Cho2000, Maron2001, 
Cho2002, BeL2006},
although some disagreement in terms 
of the measured spectral slope was noted. This 
disagreement produced a flow of papers with 
suggestions to improve the GS95 model 
by including additional effects like dynamical alignment 
\citep{Boldyrev2005, Boldyrev2006}, 
polarization intermittency \citep{BeL2006}, 
non-locality \citep{Gogoberidze2007}. 
More recent studies in \cite{BeL2009, BeL2010}
indicate that numerical 
simulations do not have sufficiently extended inertial range to get 
the actual spectral slope\footnote{\cite{BeL2010}
noticed that
the magnetic turbulence is less local compared with the hydrodynamic 
one and therefore 
one requires a substantially larger resolution to
distinguish the actual spectral slope from the slope affected by 
the bottleneck.}
and therefore worries about the ``inconsistency'' of the GS95 model are 
premature. 
Evidence of the GS95 spectrum for the MHD incompressible turbulence 
was recently obtained by \cite{Be2011}.

We shall add parenthetically that in a number of applications the empirical 
so-called 
composite 2D/slab model of magnetic fluctuations is used. 
In the latter model, which is also known as {\it two-component model}, 
it is assumed that fluctuations can be described as a 
superposition of fluctuations with wave vectors parallel to the ambient 
large-scale 
magnetic field (so-called {\it slab modes}) and perpendicular 
to the mean field (so-called 
{\it two-dimensional modes}). 
It results in a {\it maltese cross} structure of magnetic 
correlations. 
This model was developed to account for the solar wind observations, 
which it does 
well by adjusting the intensity of the two components
\citep[e.g.][]{Matthaeus1990}.
This theory of 2D fluctuations is consistent with the theory of weak 
Alfvenic turbulence \citep[e.g., ][]{Ng1996, LV1999, Galtier2000}
but it can describe Alfvenic turbulence only over a limited range of scales.
It may be treated 
as a parameterization of a particular type of magnetic perturbation 
dominated by the 
peculiarities of driving, but
recent simulations by Gosh (2011, private communication) show that, 
at best, the model 
represents a special transient
state of a not fully developed turbulence. 
In addition, slab modes do not arise naturally in turbulence with 
large-scale driving, as shown
by MHD numerical simulations \citep{ChoL2002, ChoL2003}.
Thus we do not consider this model for clusters of galaxies. 

The GS95 model of turbulence can be adopted to describe the 
Alfvenic part of MHD turbulent 
fluctuations in galaxy clusters. The model can be 
generalized also to compressible turbulence and even for supersonic motions 
numerical calculations show that the Alfvenic perturbations exhibit 
GS95 scaling \citep{ChoL2002, ChoL2003, Kowal2010}.
We note that we consider MHD turbulence where the flows of energy 
in the opposite directions are balanced. 
When this is not true, i.e. when the turbulence has non-zero 
cross-helicity, the properties of turbulence depart substantially 
from the GS95 model\footnote{Among the existing theories of imbalanced 
turbulence \citep[e.g.,][]{Lithwick2007, BeL2008, Chandran2008, Perez2009},
all, but \cite{BeL2008} contradict to numerical testing in \cite{BeL2009, 
BeL2010} }.
Solar wind presents a system with imbalanced turbulence.
However, the degree of imbalance of turbulence
in clusters of galaxies is unclear and we know that in compressible media 
the imbalance decreases due to reflecting 
of waves from pre-existing density fluctuations and 
due to the development of parametric instabililites 
\citep{DelZanna2001}.
Similarly, we shall not discuss MHD turbulence at high magnetic 
Prandtl numbers,
when the viscosity is much larger than resistivity
\citep[e.g.,][]{Cho2002, Cho2003}.

The GS95 model of turbulence combined with several considerations
on the macro- and micro-physics of the IGM allows for a basic picture of
the properties of turbulence in galaxy clusters \citep[e.g., ][]{BL2007}.

\noindent
Cosmological numerical simulations show that large-scale turbulent motions
are generated during the process of cluster formation
(\cite{Dolag2005, Iapichino2008, Vazza2011}, see also Nagai 2011, Iapichino
2011, Vazza 2011, this conference). These motions, injected at large
scales $L_o \sim 300-500$ kpc, are believed to provide
the driver for turbulence at smaller scales.
The typical velocity of the turbulent eddies at the injection scale
is expected to be
around $V_L \sim 500-700$ km/s which makes turbulence sub--sonic,
but strongly super--Alfv\'enic.
Turbulence at large scales is thus essentially hydrodynamic and 
-- most likely -- made of a mix of compressive and incompressive eddies.
The cascading of compressive (magnetosonic) modes may indeed couple with
that of solenoidal motions (Kolmogorov eddies).

\noindent
Viscosity in a turbulent and magnetised IGM is strongly suppressed due
to the effect of the bending of magnetic field lines and of the
perturbations of the magnetic field induced by plasma 
instabilities (e.g., Sect.~2.1).
The important consequence is that an inertial range in the IGM
is established -- for both solenoidal and compressive modes -- 
down to collisionless scales where a fraction of the turbulent energy
is channelled into acceleration/heating of CR and thermal plasma
(see \cite{BL2007, BL2011a} for discussion).
At small scales -- in the inertial range -- the velocity of turbulent
eddies becomes sub-Alfvenic and turbulence is described in the
MHD regime. At these scales the coupling between Alfv\'en
and compressible modes gets changed
and only slow modes are cascaded by Alfv\'enic modes 
\citep[e.g.,][]{GS1995, Lithwick2001, ChoL2002}. 
The cascading of fast modes is not particularly
sensitive to the presence of the other modes, fast modes remain
isotropic while the spectrum of other modes becomes anisotropic.

\subsection{Spectroscopic ways of turbulence studies}

Recent observational advances to constrain turbulence in the IGM focussed 
on the
broadening of lines in the X-ray spectra of galaxy clusters and provide 
interesting
limits in the case of cool-core clusters \citep[e.g.,][]{Sanders2011}.

Turbulence in clusters of galaxies can be studied in future
using Doppler broderned emission.
Here we briefly review techniques originally developed
for studies of Doppler broderned emission and absorption lines 
in the interstellar medium research.
These techniques, Velocity Channel Analysis (VCA) and 
Velocity Correlation Spectrum (VCS) have been developed by 
\cite{LP2000, LP2004, LP2006, LP2008}
(henceforth LP00, LP04, LP06, LP08, respectively) and successfully
used for studying turbulence in diffuse and molecular gas 
\citep{Lazarian2009, Padoan2009, Chepurnov2010}.
These techniques can be applied -- at some extent -- to the case
of the IGM and future 
X-ray telescopes with very good spectroscopic capabilities
(eg ASTRO-H) can be used for the studies. 

The difference between the VCA and the VCS is how the data is being handled. 

\noindent
In the VCA technique the Position-Position-Velocity data
cubes available through spectroscopic observations are analysed by taking 
spectrum of the 
velocity slice of the cube. The spectrum of
the fluctuations is analysed while changing the thickness of the velocity 
slice and the 
analytical description of the statistics of the fluctuations
in the PPV slices described in LP00 and LP04 is used to obtain both the 
spectrum of velocity 
and the spectrum of density fluctuations.

\begin{figure}
  \includegraphics[width=.5\textwidth]{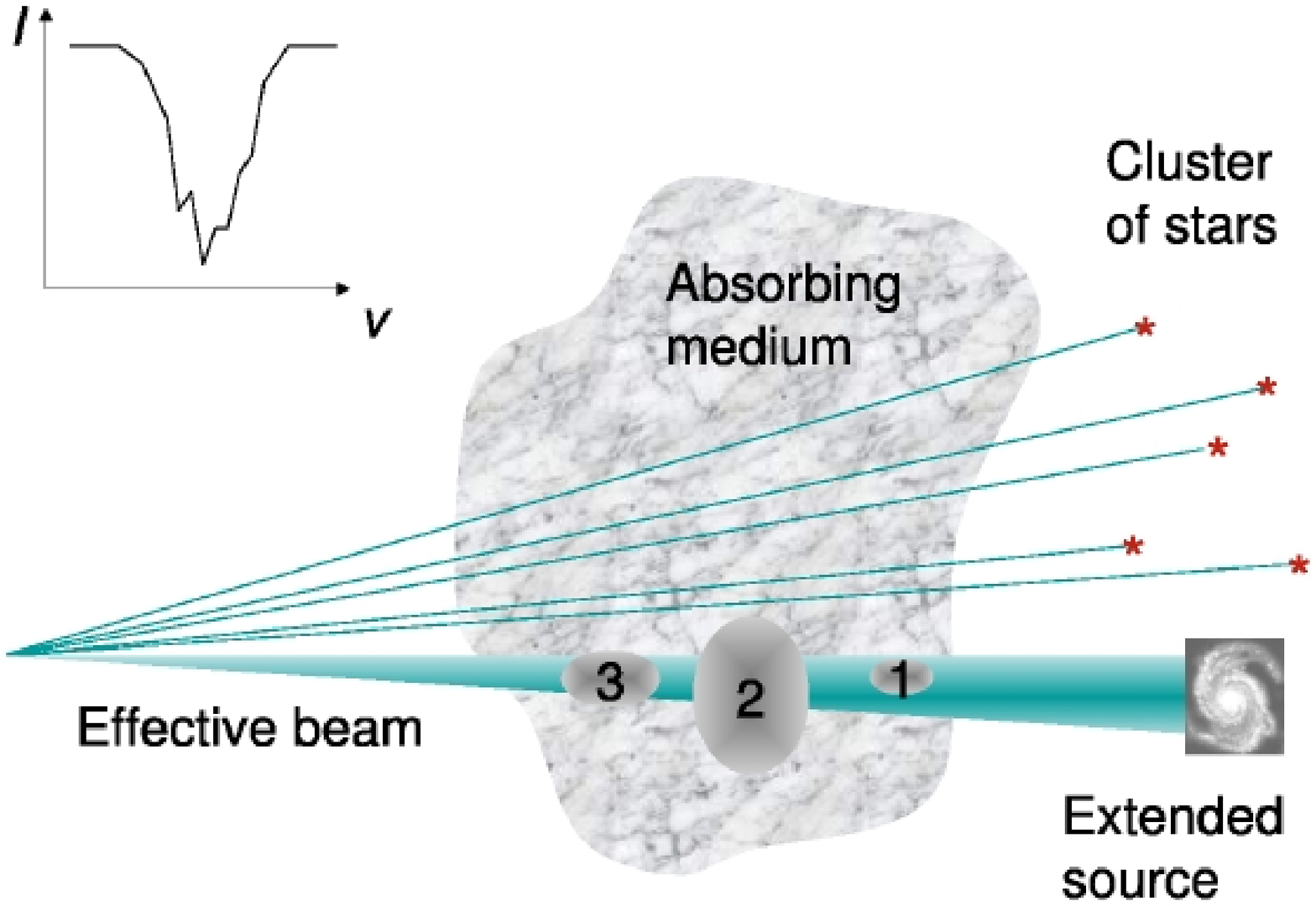}
  \includegraphics[width=.4\textwidth]{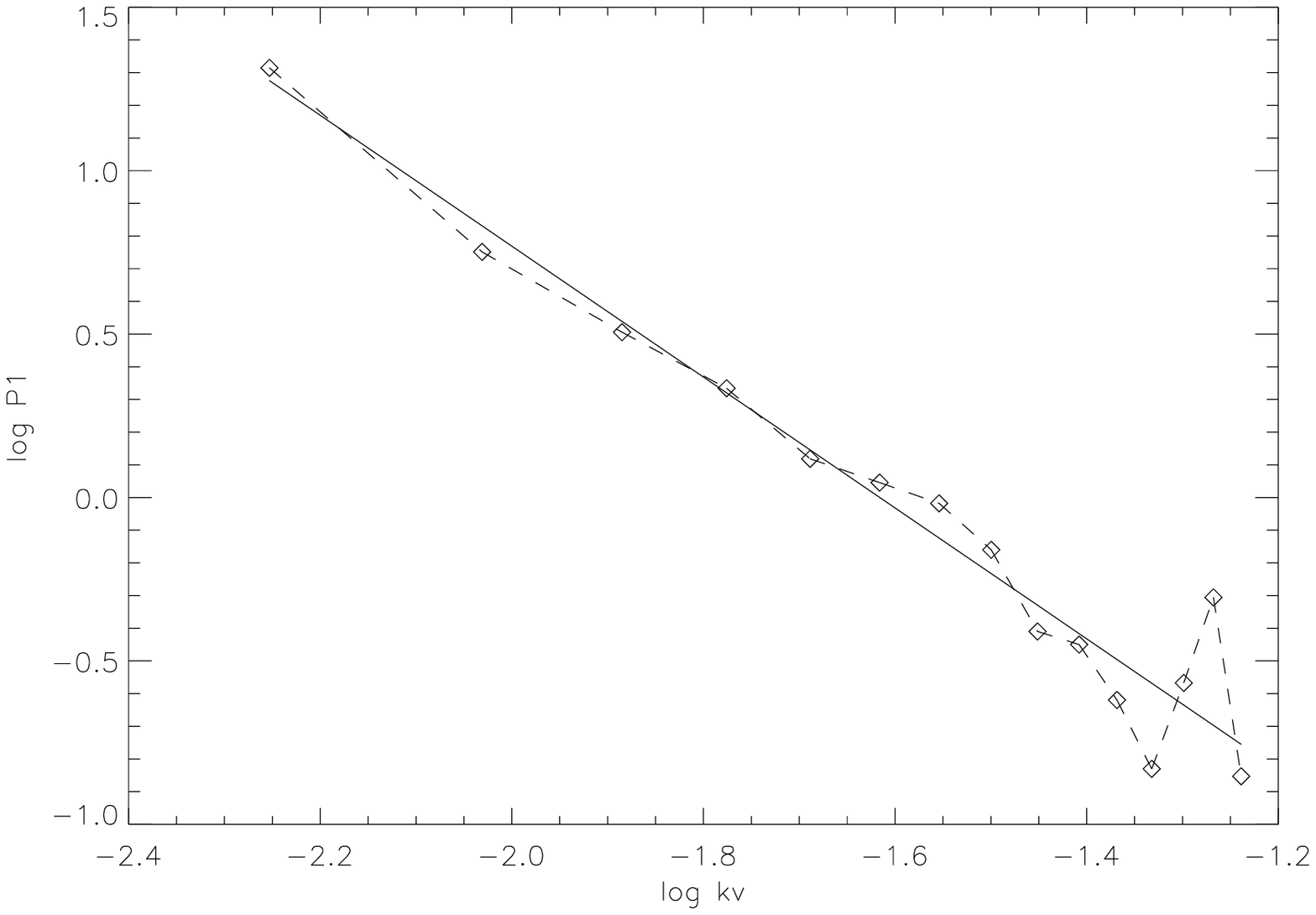}
  \caption{\footnotesize Illustration of VCS absorption studies of turbulence. {\it Upper Panel}: Schematic of measuring turbulence
  with absorption lines from point sources, e.g. stars, and an extended source, e.g. a galaxy. {\it Lower panel}: Velocity Coordinate Spectrum obtained
  using sampling of a turbulent volume along 10 lines of sight.  The solid line corresponds to the theoretical expectations. 
  Readapted from \cite{ChLazarian2009}. }
\label{abs10}
\end{figure}

\noindent
A different approach is used in the VCS technique, 
where fluctuations are analysed along 
the velocity coordinate. For the VCS technique one does not
require good coverage of the Position-Position plane and a few spectral lines 
are sufficient to get the spectra of velocity and density 
(see Figure \ref{abs10}).

New effects arise when strong absorption lines, which are in a saturated 
regime, are studied. 
The procedure for studying of the saturated lines is presented in LP08.

Our study of the effect of finite temperatures for the technique reveals that,
unlike the VCA, the temperature broadening does not prevent the turbulence
spectrum from being recovered from observations. Indeed, in VCA, gas
temperature acts in the same way as the width of a channel. Within the VCS
the term with temperature gets factorized and it influences the amplitude
of fluctuations (LP06). 
One can correct for this term\footnote{To do this, one may
attempt to fit for the temperature that would remove the exponential 
fall off in the spectrum of fluctuations along the velocity coordinate
\citep{ChL2006} }, which also allows for a new
way of estimating the interstellar gas temperature.

\noindent
Another advantage of the VCS compared to the VCA is that it reveals the
spectrum of turbulence directly, while within the VCA the slope of the spectrum
should be inferred from varying the thickness of the channel. As the thermal
line width acts in a similar way as the channel thickness, additional care
(see LP04) should be exercised not to confuse the channel that is still
thick due to thermal velocity broadening with the channel that shows the 
thin slice asymptotics. A simultaneous
use of the VCA and the VCS makes the turbulence
spectrum identification more reliable.

Both VCA and VCS are applicable to studies of not only 
emission, but also absorption lines.  
We note, that while dealing with emission lines we may face additional 
complications. 
For instance, Lazarian \& Pogosyan (see LP00, LP04, LP06, LP08) treated 
the emissivities 
proportional to the density to the first power. 
Therefore, in terms of scalings,
the emissivities and densities were interchangeable. 
This is not true, however, 
when the emissivities are proportional to $\rho^2$, as is the case of the 
recombination 
lines in plasma. The latter regime modifies the analysis. 
In particular, for the shallow spectrum of density, \cite{ChL2006}
showed that the power spectrum of density $P_{\rho}\sim k^{-\alpha}$ has 
a shallow spectral 
index $\alpha<3$ emissivity spectrum $P_{\epsilon}\sim k^{\alpha_{\epsilon}}$  
is $\alpha_{\epsilon}=2\alpha-3$ 
and this index should be used in all the expressions obtained of the 
VCA and VCS techniques. 
For the steep power law index of density, the power law indexes of the 
emissivity and density 
coinside for sufficiently large wavenumbers $k$. 

\section{Reconnection and Reconnection Diffusion}

It is generally believed that magnetic field embedded in a highly conductive 
fluid preserves its topology for all time due to magnetic fields being 
frozen-in \citep[e.g.,][]{Alfven1942, Parker1979}.
Although ionized astrophysical objects are almost perfectly 
conducting, they show indications of changes in topology, 
``magnetic reconnection'', on dynamical time 
scales \citep[e.g.,][]{Lovelace1976, Priest2000}.
Reconnection can be observed directly in the solar corona
\citep[e.g.,][]{Yokoyama1995, Masuda1994},
but can also be inferred from the 
existence of large-scale dynamo activity inside stellar interiors 
\citep[e.g.,][]{Parker1993}.
Also Solar flares \citep[][]{Sturrock1966}
and $\gamma$-ray bursts \citep[e.g.,][]{Fox2005, Galama1998}
are usually associated with magnetic reconnection. 

\begin{figure}
\hbox{
\includegraphics[width=.5\textwidth]{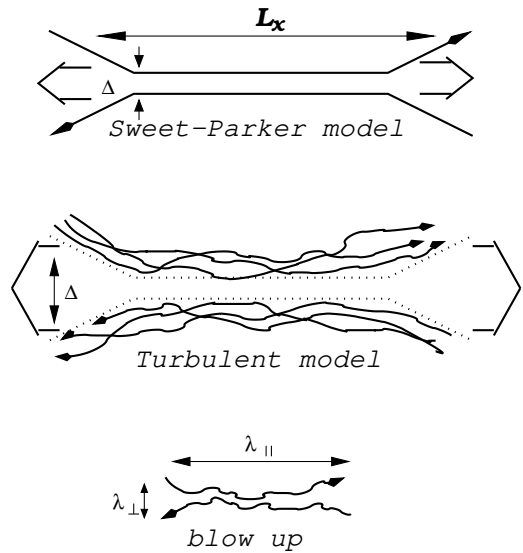}
}
\caption{\footnotesize {\it Upper plot}:
Sweet-Parker model of reconnection. The outflow
is limited by a thin slot $\Delta$, which is determined by Ohmic
diffusivity. The other scale is an astrophysical scale $L\gg \Delta$.
{\it Middle plot}: Reconnection of weakly stochastic magnetic field according to
LV99. The model accounts for the stochasticity
of magnetic field lines. The outflow is limited by the diffusion of
magnetic field lines, which depends on field line stochasticity.
{\it Low plot}: An individual small scale reconnection region. The
reconnection over small patches of magnetic field determines the local
reconnection rate. The global reconnection rate is substantially larger
as many independent patches come together. From \cite{L2004}. }
\label{LV99}
\end{figure}

To understand the difference between reconnection in astrophysical 
situations and 
in numerical simulations, one should 
recall that the dimensionless combination that
controls the reconnection rate is the Lundquist number\footnote{The magnetic
Reynolds number, 
which is the ratio of the magnetic field decay time to the eddy
turnover time, 
is defined using the injection velocity $v_l$ as a characteristic
speed instead of the Alfv\'en speed $V_A$, which is taken in the Lundquist
number.}, defined as $S = L_xV_A / \lambda$, where $L_x$ is the length of the
reconnection layer, 
$V_A$ is the Alfv\'en velocity, and $\lambda=\eta c^2/4\pi$ 
is Ohmic diffusivity. Because of the huge astrophysical 
length-scales $L_x$ involved, 
the astrophysical Lundquist numbers are also huge,
e.g. for the IGM they can be as high as $10^{20}$, 
while present-day MHD simulations
correspond to $S<10^4$. 
As the numerical efforts scale as $L_x^4$, where $L_x$
is the size of the box, 
it is feasible neither at present nor in the foreseeable future 
to have simulations with sufficiently high Lundquist numbers.

Observations have always been suggestive 
that magnetic reconnection can happen at high speed, 
in spite of theoretical difficulties to explain the effect. At the same, 
the phenomenon of solar flares was suggestive 
that magnetic reconnection may be slow in order 
to ensure the accumulation of magnetic flux and suddenly gets fast to 
explain the observed fast release of energy. 
A model that can naturally explain this and other 
observational manifestations of magnetic reconnection was proposed in 
\cite{LV1999} (LV99).
The model appeals to the ubiquitous astrophysical turbulence as a universal 
trigger and controller of fast reconnection. 

To deal with strong, dynamically important magnetic fields, LV99 proposed 
a model of fast reconnection in the presence of 
sub-Alfv\'enic turbulence (see Figure~\ref{LV99}). 
They identified stochastic wandering of magnetic field-lines 
as the most critical property of MHD turbulence which permits fast 
reconnection.  
As we discuss more below, this line-wandering widens the outflow region 
and alleviates the controlling constraint of mass conservation. The LV99 model 
has been successfully tested recently in \cite{Kowal2009}
(see also higher resolution results in \cite{L2010}).
The model is radically different from its predecessors which also appealed 
to the effects of turbulence. For instance, unlike \cite{Speiser1970}
and \cite{Jacobson1984} 
the model does not appeal to changes of microscopic properties of 
plasma\footnote{The nearest progenitor to LV99 was the work of  
\cite{Matthaeus1985, Matthaeus1986}, who studied the problem numerically 
in 2D MHD and who suggested that magnetic reconnection 
may be fast due to a number of  turbulence effects, 
e.g. multiple X points and turbulent EMF. 
However, \cite{Matthaeus1985, Matthaeus1986} did not observe 
the important role of magnetic field-line wandering, and did not obtain 
a quantitative prediction for the reconnection rate, as did LV99.}.

The LV99 model justifies the notion of turbulent mixing perpendicular 
to magnetic field lines. Indeed,
LV99 showed that the GS95 model gets self-consistent only in the presence 
of the turbulence-induced
reconnection with the rates predicted in LV99. 
Otherwise, the formation of the magnetic knots would
change the character of the turbulent interactions.

The understanding of fast magnetic reconnection in the presence of turbulence 
induced the notion
of ``reconnection diffusion'' that was described in \cite{L2005} and later 
used for describing different phenomena from star formation to heating of 
magnetic filaments in IGM \citep[e.g., ][]{Santos2010, L2010}. 
The same concept was implicitly used earlier in \cite{Cho2003}
where it was claimed that the heat conductivity of the IGM is influenced
by the heat advection by turbulent eddies. 
Explicit calculations done by \cite{L2006}
show that the heat conduction by turbulent eddies mixing magnetic field 
perpendicular to the local direction
of magnetic field is the dominant way of heat transport in clusters 
of galaxies.
The effect of reduced
mean free path of thermal electrons induced by turbulence that we discussed 
above (Sect. 2) increases
the relative importance of thermal transfer via reconnection diffusion. 
Rigorous arguments justifying the concept of reconnection diffusion
can be found in \cite{Eyink2011}.

\section{Cosmic ray acceleration}

Radio observations of galaxy clusters probe particle acceleration by
shocks and turbulence in the IGM (Brunetti 2011, this conference for review 
on physics of cosmic rays (CR) in the IGM).
In this Section we briefly discuss the importance of turbulence in the 
acceleration of CR and the connected issue of CR acceleration induced
by magnetic reconnection.

\subsection{Acceleration by magnetic turbulence}

The interaction of turbulence and cosmic rays is a vital component of models 
of CR propagation and acceleration. It has been a concern from the very 
beginning of CR research \citep[e.g.,][]{Ginzburg1966, 
Jokipii1966, Wentzel1969}.
To account for the interaction properly, one must know both the scaling of
turbulence,
the changes with time of turbulence spectrum due to
the damping processes (e.g. with CR),
and the interactions of turbulence with various waves produced by CRs. 

Clusters of galaxies present magnetic fields of the largest extend and 
they are 
also considered on the role of the accelerators of the ultra high energy CR.
The acceleration of particles in large (Mpc) regions in
galaxy clusters is generally believed to happen via the second 
order Fermi process as a result of the interaction of particle--turbulence 
interactions \citep[e.g.,][]{BL2007, Petrosian2008, Brunetti2008}.
Similarly, acceleration by magnetic turbulence is a very robust process that 
is likely be important for Solar flares, gamma ray bursts and many 
other astrophysical environments \citep[e.g.,][]{Hamilton1992, Miller1996, 
Schlickeiser2000, Dermer2001}.

MHD turbulence is the most important for the acceleration of particles 
of largest energies and it is vital to use the theoretically justified 
and numerically tested relations in the studies of particle acceleration. 
From the start of the work in this direction \citep[e.g., ][]{Chandran2000}
it became clear that the earlier models for the acceleration and
propagation of energetic particles that were based on ad hoc representation 
of turbulence are in error of many orders of magnitude as far as Alfvenic 
perturbations are concerned. \cite{YanL2002, YanL2004}
identified compressible fast modes as the principal agent for CR
acceleration 
by MHD turbulence. As the aforementioned modes, unlike Alfvenic ones, 
are subject to rather strong damping, the description of the acceleration 
gets more complicated.
In \cite{BL2007}
we derived a comprehensive picture of compressible turbulence in
galaxy clusters and studies CR acceleration considering all the
relevant damping processes, with the results providing good correspondence
with observations.
More recently we extended this formalism to the case of the reacceleration
of CR and of the secondary particles generated in the IGM via pp collisions
\citep{BL2011b}.

\noindent
In addition, the accuracy of the particle acceleration using analytical theory 
has been improved by extending the quasi-linear theory to the regime of 
substantial perturbations of magnetic field 
and applied to the case of Solar flares \citep{YanL2008, Yan2008}.
The improved theory has been successfully tested with direct tracing of CR 
trajectories in data cubes obtained with results of direct MHD simulations 
of turbulence \citep{Beresnyak2011}.
Future applications of these extensions to the case of galaxy clusters will
be important.

\noindent
Compressible turbulence interacts both with CR and with thermal
particles. This interaction may also induce magnetic field perturbations
(trough plasma instabilities, e.g. Sect.2) 
that may further come into play in the
particle acceleration process.
First attempts in this direction suggest that the fraction of turbulence
that goes into CR acceleration increases when 
turbulent-induced instabilities are taken into account \citep{BL2011a}.

\subsection{Shock acceleration and turbulence}

Here we focus on the importance of turbulence in shock
acceleration mechanisms.
Shock acceleration is thought to be one of the principal accepted 
mechanisms of energetic particle acceleration. The shock induces compression 
and particles trapped between magnetic fluctuations ahead and behind shocks 
fill the acceleration every time they bounce back and forth between 
converging fluctuations. This is an efficient way of accelerating particles 
which results in the energy gain per bouncing to increase as the first power 
of the ratio of the particle velocity to that of light, i.e. $v/c$, making 
this process known as the first order Fermi acceleration.

\noindent
Shock acceleration in galaxy clusters is believed to contribute
the most of the CR (protons), while shock acceleration of CR electrons
is the most popular model to explain the origin of radio relics
(\cite{Ensslin1998}, Ryu 2011, Br\"uggen 2011,
this conference for review).

The necessity of particles to bounce back and forth limits the efficiency 
of the acceleration of high energy particles through a requirement 
that the energetic particle  should have the Larmor radius less than 
size of the magnetic fluctuations that they bounce off. 
Therefore to increase the energy of the accelerated particles one should 
have strong magnetic field and strong magnetic fluctuations both in the 
preshock and postshock regions. The situation with the postshock region 
is relatively simple. Gas passing through the shocks is known to create 
turbulence \citep[e.g.,][]{Giacalone2007}.
The turbulence is known to increase the magnetic field energy, enabling 
particles to scatter efficiently and return to the shock region for 
further acceleration. For the preshock region, most work was concentrated 
on instabilities that can enhance magnetic field. The most commonly discussed 
is the so-called Bell instability \citep{Bell2004} which is a non-resonant 
current driven instability, that can 
increase magnetic field in front of the shock. 
In \cite{BeJL2009} we proposed that a turbulent generation of magnetic 
field is happening in 
front of the shock, in the region which is called precursor. The properties 
of precursor and its formation in front of the shock are described in the 
literature \citep[e.g.,][]{Malkov2001}. 
As the precursor interacts with the density inhomogeneities 
preexisting in the medium in front of the shock, it gets perturbed, 
creating vorticity and turbulence. New studies of turbulent amplification 
of magnetic field \citep[e.g., ][]{Cho2009} 
provide the rates of magnetic field amplification by turbulence. 
These rates were made use of in \cite{BeJL2009}
to obtain the values of the turbulent magnetic field that is 
generated in front of the shock. The corresponding estimates show that 
the preshock magnetic fields produced via this process are larger than those 
arising from the Bell instability and that they 
account for cosmic ray acceleration in galactic supernovae shock up 
to $10^{15}$~eV, the so-called "knee" of the cosmic ray spectrum. 

Further development of this direction presents a very promising avenue 
of the cosmic ray acceleration research. Interesting boot strap processes 
are likely to be at work as generation of magnetic 
fluctuations in front of the shock increases the 
efficiency of the acceleration, 
contributing to the development of the precursor.

\subsection{Acceleration induced by magnetic reconnection}

\begin{figure}
\hbox{
\includegraphics[width=5.0 cm]{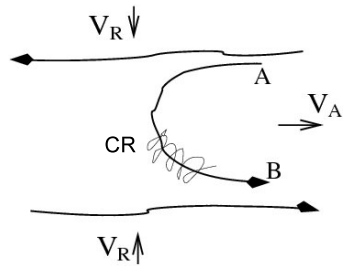}
}
\caption{\footnotesize CR spiral about a reconnected magnetic 
field line and bounce back at points A and B. The reconnected regions 
move towards each other with the reconnection velocity $V_R$. 
The advection of cosmic rays entrained on magnetic field lines happens 
at the outflow velocity, which is in most cases of the order of $V_A$. 
Bouncing at points A and B happens because either of streaming instability 
induced by energetic particles or magnetic turbulence in the reconnection 
region. In reality, the outflow region gets filled in by the oppositely 
moving tubes of reconnected flux which collide only to repeat on a smaller 
scale the pattern of the larger scale reconnection. 
Thus our cartoon also illustrates the particle acceleration taking place 
at smaller scales.} 
\label{f5}
\end{figure}

An important consequence of fast reconnection of turbulent magnetic fields 
that we discussed in Sect. 3 is the formation of a thick volume filled 
with reconnected magnetic flux loops. These 3D loops contract, presenting 
favorable conditions for energetic particle acceleration. 
This process of first order Fermi acceleration of energetic particles 
in reconnection regions has been described in \cite{deGouveia2005} 
(see also Fig. \ref{f5}) for the situation when 
there is no back reaction of the accelerated particles on the reconnected 
magnetic flux. 
\cite{Drake2006} appealed to a similar process within their preferred model of 
collisionless reconnection and proposed that firehose instability can 
play a role of the feedback for the accelerated particles.

More recently, the acceleration in reconnection regions has obtained 
observational support. 
It was suggested in \cite{LazarianO2009}
that anomalous CR measured by Voyagers are, in fact accelerated in 
the reconnection regions of magnetopose (see also \cite{Drake2010}). 
Such a model explains why Voyagers did not see any signatures 
of acceleration passing the Solar system termination shock. 
In a separate development, \cite{LazarianD2010} 
appealed to the energetic particle acceleration in the wake 
produced as the Solar system moves through interstellar gas to explain 
the excess of cosmic rays of the range of both sub-Tev and 
multi-TeV energies in the direction of the magnetotail Magnetic 
reconnection is ubiquitous in astrophysical circumstances and therefore 
it is expected to induce acceleration of particles in a wide range of 
astrophysical environments.
For instance, the process has been already discussed for the acceleration 
of particles in gamma ray bursts \citep{L2003, Zhang2011} and
microquasars \citep{deGouveia2005}.
We expect the process to be important for the acceleration of protons 
and electrons in galaxy clusters.

Numerical 2D simulations presented in \cite{Drake2010}
confirmed high efficiency of particle acceleration in regions of magnetic 
reconnection. However, results in \cite{L2010} 
show that the process of acceleration happens rather differently in 2D 
and 3D situations. The 3D geometry shows a wider variety of acceleration 
regimes and this calls for much more detailed studies of the acceleration.

\section{Summary}

The main points of our review can be summarized as follows

\begin{itemize}

\item
Turbulence is essential for understanding of the IGM.
On large scale the description of turbulence obtained in MHD can be used. 
Compressions induced by turbulence 
induce instabilities in the IGM, 
changing the mean free path of thermal ions. 
This should extend the range over which the MHD description 
of turbulence is applicable.

\item
Studies of turbulence in the IGM
can get a boost if Doppler-broaderned 
spectral emission and absorption lines are used. The techniques
originally developed and successfully used in the interstellar 
research, namely Velocity Channel Analysis (VCA) and 
Velocity Correlation Spectrum (VCS)
are promissing for studing of turbulence in the IGM.

\item
Magnetic reconnection happens fast in turbulent media, which makes 
the models of MHD turbulence self-consistent. Fast magnetic reconnection
makes MHD turbulence somewhat similar to hydrodynamic if one 
considers turbulent motions perpendicular to the local 
direction of magnetic field. 
Such motions can induce a process 
of ``reconnection diffusion'' which efficient heat transfer in the IGM.

\item
Magnetic turbulence is very important for particle acceleration in 
clusters of galaxies. 
It can accelerate particles through direct interactions with 
turbulent fluctuations. However, it can also modify shocks, inducing magnetic 
field generation in shock precursors and increasing the efficiency of
high energy particle acceleration by shocks. In addition, it can enable 
fast magnetic reconnection which can accelerate particles within the thick
reconnection regions.
\end{itemize}

\begin{acknowledgements}
AL thanks the NSF grant AST 0808118, 
NASA grant NNX09AH78G and the support of the Center for Magnetic 
Self Organization.
GB acknowledge partial support from PRIN-INAF 2008, 2009.
\end{acknowledgements}

\bibliographystyle{aa}

\end{document}